\documentclass[twocolumn,amsmath,showpacs,aps,prb,10pt]{revtex4-1}

\usepackage{graphicx}
\usepackage{color}
\usepackage{times}

\begin{document}

\title{Probing barrier transmission in ballistic graphene}

\author{Daniel~Gunlycke}
\email{daniel.gunlycke@nrl.navy.mil}
\affiliation{Naval Research Laboratory, Washington, D.C. 20375, USA}

\author{Carter~T.~White}
\affiliation{Naval Research Laboratory, Washington, D.C. 20375, USA}

\begin{abstract}
We derive the local density of states from itinerant and boundary states around transport barriers and edges in graphene and show that the itinerant states lead to mesoscale undulations that could be used to probe their scattering properties in equilibrium without the need for lateral transport measurements.  This finding will facilitate vetting of extended structural defects such as grain boundaries or line defects as transport barriers for switchable graphene resonant tunneling transistors.  We also show that barriers could exhibit double minima and that the charge density away from highly reflective barriers and edges scales as $x^{-2}$.
\end{abstract}



\maketitle


One challenge currently preventing widespread use of graphene in nanoelectronic devices is the absence of a band gap at the Fermi level.  Without a practical band gap, other ways to switch on and off electron and hole currents are needed.  A promising possibility is to use graphene transport barriers formed by extended structural defects such as grain boundaries \cite{Malo10,Yazy10,Yazy10a,Huan11,Yu11,Kim11,Tapa12,Ahma12,Koep12,Clar13,Tiso14} or line defects.\cite{Appe10,Lahi10,Gunl11,Chen14}  It has been shown that two such barriers in a parallel configuration produces a graphene resonant tunneling transistor with an appreciable transport gap and perfect valley filtering.\cite{Gunl13}  This approach, however, requires transport barriers that are both penetrable and fairly reflective.  In vetting potential candidates, it would be advantageous to be able to probe the barrier transmissivity without having to perform lateral transport measurements.

\begin{figure}[h!]
    \includegraphics{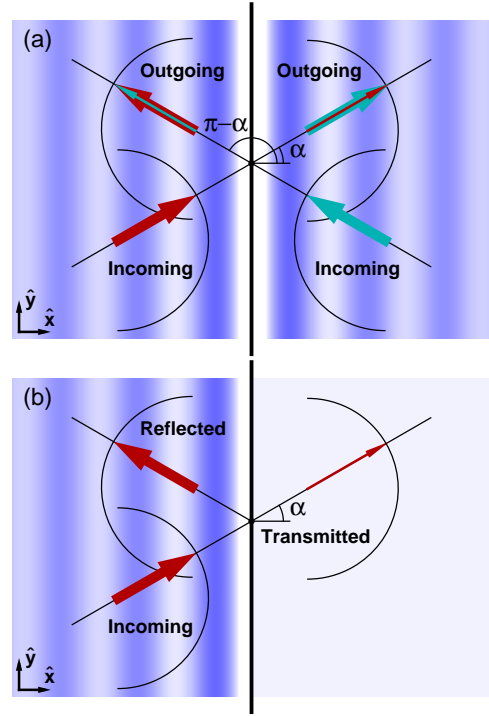}
    \caption{Schematic illustrations of the relationship between the barrier transmissivity and the undulations in the LDOS itinerant from (a) both sides and (b) the left side, representing equilibrium and non-equilibrium, respectively.  Because carriers originating from the left and right do not interfere, the undulations are the same in (a) and (b), except for the transmitted side in (b), where they are absent.  The transmission probability through the barrier is related to the undulations, which could be probed in equilibrium, even through no net current flows through the barrier.
}
    \label{f.1}
\end{figure}

While it might seem impossible to probe the conduction through a barrier with no current, we show herein that the quantum nature of the charge carriers let us do exactly that.  We combine two properties enabled by the wave-particle duality: quantum tunneling and quantum interference.  Quantum tunneling allows carriers to transmit across narrow barriers as evanescent waves.  Line defects and grain boundaries with a limited transmission probability are examples of such barriers in graphene.\cite{Gunl13,Garg14}  As illustrated in Fig.~\ref{f.1}(a), these barriers in equilibrium are surrounded by density undulations.  These mesoscale undulations are related to Friedel oscillations from isolated impurity sites \cite{Frie52,Chei06_2,Chei07_2,Hwan08,Bacs10} and arise because the limited number of wave vectors allowed by the band structure is unable to describe sharp real-space features.  We show that the undulations result from quantum interference between incoming and outgoing waves, but not all outgoing waves---and this is the key---only the reflected waves.  The undulations on the left side of the barrier in Fig.\,\ref{f.1}(a) are therefore the same as those in Fig.\,\ref{f.1}(b), which illustrates non-equilibrium with carriers originating from the left side only.  This is the connection that allows us to probe the transmission probability through the barrier, even in equilibrium.

To understand the relationship between the transmissivity and the undulations in the local density of states (LDOS) in equilibrium, we express the LDOS at energy $E$ and coordinate $x$ ({\it cf.} Fig.\,\ref{f.1}), centered at the barrier, as
\begin{equation}
	\rho_E^{}(x) = \rho_E^\rightarrow(x) + \rho_E^\leftarrow(x) + \rho_E^\mathrm{(b)}(x),
	\label{e.1}
\end{equation}
where the terms, respectively, represent the local densities of states itinerant from the left and right sides of the barrier, and boundary states at the barrier.

It can be shown that the LDOS per unit area originating from the left side is
\begin{equation}
	\rho_E^\rightarrow(x)=\frac{\rho_E^{}}{4\pi}\sum_\tau\int_{-\pi/2}^{\pi/2}\operatorname{Tr}\,\langle \vec r|\Psi_{\eta\tau\vec{q}}\rangle\langle\Psi_{\eta\tau\vec{q}}|\vec r\rangle\mathrm{d}\alpha,
	\label{e.2}
\end{equation}
where $\rho_E^{}\equiv\frac{2|E|}{\pi(\hbar v_F)^2}$ is the graphene density of states per unit area with the graphene Fermi velocity $v_F\approx8.5\times10^5$\,ms$^{-1}$, and $\langle\vec r|\Psi_{\eta\tau\vec{q}}\rangle$ is the wave function in the presence of the barrier.  The wave function indices are carrier type $\eta=\pm1$ for electrons and holes, respectively, valley index $\tau=\pm1$ representing the two inequivalent graphene symmetry points $K$ and $K'$, and wave vector $\vec q=q\hat q$ centered at the considered symmetry point.  For elastic scattering, we have $E=\varepsilon_{\eta\tau\vec{q}}$, where $\varepsilon_{\eta\tau\vec{q}}=\eta\hbar v_Fq$ assuming a linear graphene dispersion.  Because the carrier group velocity $\vec{v}\equiv(\eta/\hbar)\nabla_{\vec{q}}\varepsilon_{\eta\tau q}=v_F\hat{q}$, the wave vector is locked to the propagation angle $\alpha$, defined in Fig.\,\ref{f.1}, so that $\vec q=q(\hat x\cos\alpha+\hat y\sin\alpha)$.  The scattering is specular near the Dirac point, owing to energy and wave vector conservation along the barrier, and for the majority of extended structural defects, these conservation laws also prohibit intervalley scattering.\cite{Yazy10a,Gunl14}  We can express the wave function for a state originating from the left as
\begin{subequations}
	\begin{align}
		\langle\vec r|\Psi_{\eta\tau\vec{q}}\rangle &= \frac{1}{\sqrt{2}}\left(\begin{array}{c}1\\i\eta e^{i\tau\alpha}\end{array}\right)e^{iq_xx}e^{iq_yy}\nonumber\\
		&+\frac{r_{\eta\tau\alpha}}{\sqrt{2}}\left(\begin{array}{c}1\\-i\eta e^{-i\tau\alpha}\end{array}\right)e^{-iq_xx}e^{iq_yy},\\\nonumber\\
		\langle\vec r|\Psi_{\eta\tau\vec{q}}\rangle &= \frac{t_{\eta\tau\alpha}}{\sqrt{2}}\left(\begin{array}{c}1\\i\eta e^{i\tau\alpha}\end{array}\right)e^{iq_xx}e^{iq_yy},
	\end{align}
	\label{e.3}
\end{subequations}
on the left and right sides, respectively, where $r_{\eta\tau\alpha}$ and $t_{\eta\tau\alpha}$ are the reflection and transmission amplitudes, and phases associated with the microscopic structure has been dropped, for clarity.  Inserting Eq.\,(\ref{e.3}) into Eq.\,(\ref{e.2}) yields
\begin{equation}
	\rho_E^\rightarrow(x) = \left\{
	\begin{array}{lr}
		\Big.\left(\rho_E^{}/2\right)\left(1+\bar{R}\right)+\Delta\rho_E^{}(x), & \quad(x<0),\\
		\Big.\left(\rho_E^{}/2\right)\bar{T}, & \quad(x>0),
	\end{array}
	\right.
	\label{e.4}
\end{equation}
where $\bar{R}$ and $\bar{T}$ are the reflection and transmission probabilities, $R=|r|^2$ and $T=|t|^2$, respectively, averaged over all angles of incidences, and 
\begin{equation}
	\Delta\rho_E^{}(x)=\frac{\rho_E^{}}{4\pi}\sum_\tau\int_{-\pi/2}^{\pi/2}\operatorname{Re}\Delta_{\eta\tau\alpha}\mathrm{d}\alpha,
	\label{e.5}
\end{equation}
describes the undulations resulting from the quantum interference term $\operatorname{Re}\Delta_{\eta\tau\alpha}$ on the incoming and reflected side, where $\Delta_{\eta\tau\alpha}\equiv r_{\eta\tau\alpha}\left(1-e^{-2i\tau\alpha}\right)e^{2iq|x|\cos\tau\alpha}$.  No such undulations are present on the transmitted side, where $\rho_E^\rightarrow(x)$ is proportional to the average transmission probability.

\begin{figure}
	\includegraphics{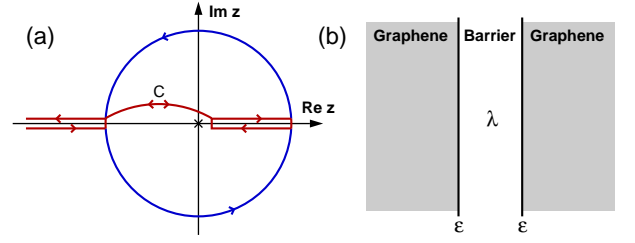}
	\caption{Contours to determine LDOS around a graphene barrier.  (a) The unit circle and real axis contours account for the itinerant and boundary states in graphene, respectively.  (b) These contributions are evaluated for a barrier with an effective coupling $\lambda$ and effective potential $\varepsilon$.}
	\label{f.2}
\end{figure}

Let us now take advantage of present symmetry.  First, we apply the parity operator $\mathcal{P}_x$ to $\rho_E^\rightarrow(x)$, which leads to
\begin{equation}
	\rho_E^\leftarrow(x)=\rho_E^\rightarrow(-x).
	\label{e.6}
\end{equation}
Next, we note that both the valley index $\tau$ and the propagation angle $\alpha$ are odd under the parity operator $\mathcal{P}_y$.  The reflection amplitude, on the other hand, is even and can be shown to only depend on $\tau$ and $\alpha$ through the product $\tau\alpha$.  Because $\Delta_{\eta\tau\alpha}$ is also even, we conclude, as one might expect, that the two valleys contribute the same amount to the undulations.  Applying the time-reversal operator $\mathcal{T}$ to the wave function $\langle\vec r|\Psi_{\eta\tau\vec{q}}\rangle$ yields $r_{\eta,-\tau,\pi+\alpha}=r^*_{\eta\tau\alpha}$, and concomitantly $\Delta_{\eta,-\tau,\pi+\alpha}=\Delta^*_{\eta\tau\alpha}$.  This allows the integral in Eq.\,(\ref{e.5}) to be formulated as a contour integral using $z\equiv ie^{i\tau\alpha}$.  Identifying the generating function for Bessel functions of the first kind and expressing it as a Laurent series lets us express the undulation term as
\begin{equation}
	\Delta\rho_E^{}(x)=\frac{\rho_E^{}}{2}\sum_{n=-\infty}^\infty I_n[r]\,J_n\left(2q|x|\right),
	\label{e.7}
\end{equation}
where the Bessel function coefficients
\begin{equation}
	I_n[r]\equiv\frac{1}{2\pi i}\oint_\mathrm{u.c.} r_\eta(z)\left(z^{n-1}+z^{n-3}\right)\mathrm{d}z
	\label{e.8}
\end{equation}
are functionals of the reflection amplitude.  This functional dependence on the reflection amplitude establishes the connection between the scattering properties and the LDOS that enables the transmission probability to be probed in equilibrium.  Poles within the reflection amplitude could also lead to LDOS contributions from boundary states.  These contributions could be expressed as
\begin{equation}
	\rho_E^\mathrm{(b)}(x)=\frac{\rho_E^{}}{4\pi i}\int_\mathcal{C} r_\eta(z)\left(z^{-1}+z^{-3}\right)e^{q|x|(z-z^{-1})}\mathrm{d}z,
	\label{e.9}
\end{equation}
where the contour $\mathcal{C}$ is shown in Fig.\,\ref{f.2}(a), and have a localization length $\xi\equiv\max_{z_p} [q(z_p-z_p^{-1})]^{-1}$ with $z_p$ being poles.

Summing everything up, we obtain the equilibrium LDOS
\begin{equation}
	\rho_E^{}(x)=\rho_E^{}+\Delta\rho_E(x)+\rho_E^\mathrm{(b)}(x)
	\label{e.10}
\end{equation}
from Eq.\,(\ref{e.1}), where we have used $\bar R+\bar T=1$, required by carrier conservation.  The charge density has the same form
\begin{equation}
	n_\mu^{}(x)=n_\mu^{}+\Delta n_\mu(x)+n_\mu^\mathrm{(b)}(x),
	\label{e.11}
\end{equation}
where each term is $-e$ times the integral over the corresponding term in Eq.\,(\ref{e.10}) from the Dirac point $E=0$ to the electrochemical potential $E=\mu$.

\begin{figure}
    \includegraphics{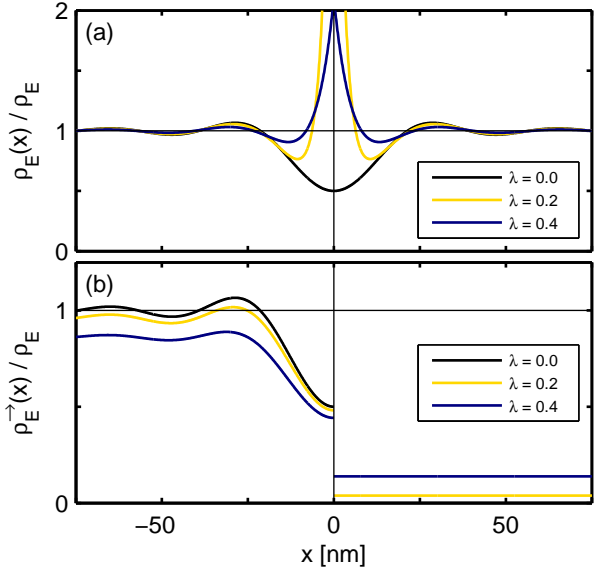}
        \caption{Local densities of states around neutral barriers ($\varepsilon=0$) at the carrier energy $E=50$\,meV that include (a) all contributions and (b) contributions from states originating from the left.  Contributions from transmitted carriers in (b) are proportional to the average transmission probability and do not exhibit undulations.}
    \label{f.3}
\end{figure}
The expressions above follow from from the properties of graphene and the barrier geometry.  They can be used with different barrier models.\cite{Gunl11,Rodr12,Eber14,Gunl14}  For illustration, we adopt below a model \cite{Gunl14} that consists of an effective coupling parameter across the barrier $\lambda$ and an effective barrier potential parameter $\varepsilon$ [{\it cf.} Fig.\,\ref{f.2}(b)], both in units of the effective graphene nearest-neighbor hopping parameter $\gamma=-2\hbar v_F/\sqrt{3}a$ with $a\approx0.246$\,nm being the graphene lattice constant.  The reflection and transmission amplitudes are then given by
\begin{subequations}
	\begin{align}
		r_\eta(z) &\equiv -1+\frac{1}{2}\left[\frac{1-z^2}{1+(\eta\varepsilon+\lambda)z}+\frac{1-z^2}{1+(\eta\varepsilon-\lambda)z}\right],\\	
		t_\eta(z) &\equiv \frac{1}{2z}\left[\frac{1-z^2}{1+(\eta\varepsilon+\lambda)z}-\frac{1-z^2}{1+(\eta\varepsilon-\lambda)z}\right],
	\end{align}
	\label{e.12}
\end{subequations}
respectively.  From the reflection amplitude, we find that the undulation term for this barrier model is
\begin{equation}
	\Delta\rho_E^{}(x)=-\frac{\rho_E^{}}{2}\left[J_0+J_2-\sum_{n=1}^\infty\frac{z_+^n+z_-^n}{2}(J_{n-2}-J_{n+2})\right],
	\label{e.13}
\end{equation}
where the Bessel function argument $2q|x|$ is implicit and
\begin{equation}
	z_\pm = \left\{
	\begin{array}{cc}
		\Big.\big(\eta\varepsilon\pm\lambda\big), & \qquad|\eta\varepsilon\pm\lambda|\le1,\\
		\Big.-\big(\eta\varepsilon\pm\lambda\big)^{-1}, & \qquad|\eta\varepsilon\pm\lambda|>1.\\
	\end{array}
	\right.
	\label{e.14}
\end{equation}
The LDOS contribution from the boundary states is
\begin{equation}
	\rho_E^\mathrm{(b)}(x)=\frac{\rho_E^{}}{4}\sum_{\forall z_p\in\{z_\pm>0\}}\left(z_p^{-2}-z_p^2\right)e^{q|x|(z_p-z_p^{-1})}.
	\label{e.15}
\end{equation}

Further insight into the density undulations can be obtained from the two barrier limits: neutral barriers ($\varepsilon=0$) and decoupled barriers ($\lambda=0$).

\begin{figure}
    \includegraphics{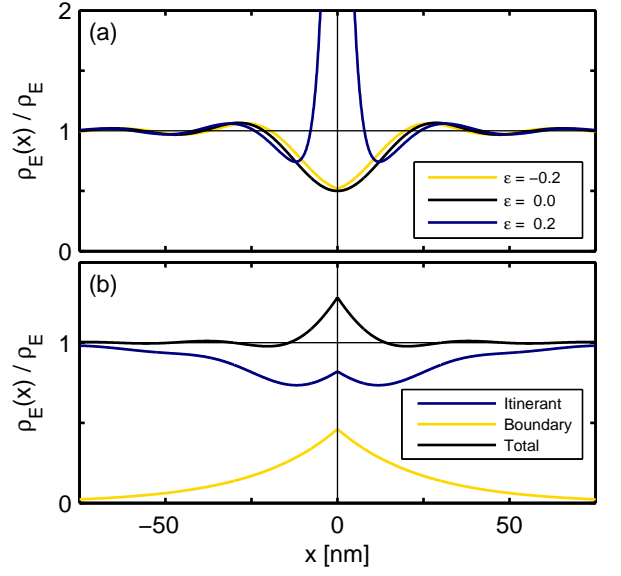}
    \caption{Local densities of states around decoupled barriers ($\lambda=0$) at the carrier energy $E=50$\,meV.  The effective potential shifts the undulation peaks in (a), and if positive, leads to states localized around the barrier.  The boundary states take their measure from the itinerant states, as shown in (b) for $\varepsilon=0.8$.}
    \label{f.4}
\end{figure}
{\it Neutral barriers} ($\varepsilon=0$):  These barriers have smooth undulations, owing to the cancellation of all odd order Bessel functions.  This smoothness, however, is concealed by the presence of a boundary state, except in the weak coupling limit $|\lambda|\rightarrow0$.  In this limit, the boundary state becomes restricted to $E=0$ and $\Delta\rho_E^{}(x)\rightarrow-(\rho_E^{}/2)(J_0+J_2)$ as $|\lambda|\rightarrow0$, as previously found for a neutral graphene edge.\cite{Fuji96_1,Jask11,Sasa10}  As $|\lambda|$ increases, the undulations fade, as shown in Fig.\,\ref{f.3}, and vanish, $\Delta\rho_E^{}(x)\rightarrow0$, in the limit $|\lambda|\rightarrow1$, where there is no barrier, and hence no quantum interference.

{\it Decoupled barriers} ($\lambda=0$): This case also describes isolated graphene edges.  The undulations are generally not smooth at the barrier and the undulation peaks shift as a function of the potential, as can be seen in Fig.\,\ref{f.4}(a).  As $\eta\varepsilon\rightarrow0$, we approach the case of neutral decoupled barriers or edges mentioned above.  We additionally find that the charge density undulations for these barriers or edges are given by 
\begin{equation}
	\Delta n_\mu^{}(x)=-\frac{n_\mu^{}}{2q_\mu^2x^2}\left[1-J_0(2q_\mu|x|)\right].
	\label{e.16}
\end{equation}
The predicted quadratic decay of the charge density undulations away from the edge is different from the cubic decay of the Friedel oscillations away from isolated impurity sites.\cite{Chei06_2}

\begin{figure}
	\vskip 0.08 in 
	\includegraphics{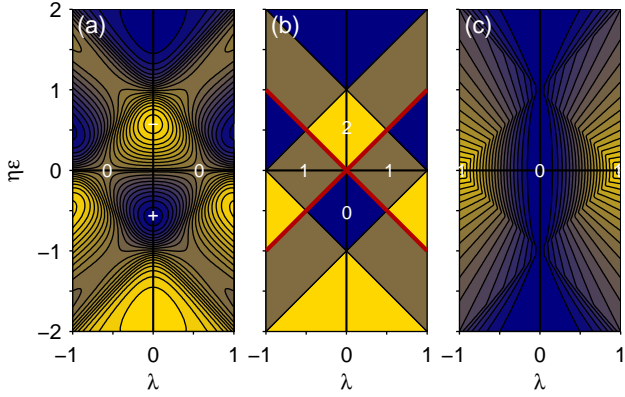}
	\caption{Properties in parameter space.  The derivative of the density undulations (a) at the barrier $\Delta\rho_E^{\prime}(0^+)$ correlates with the number of boundary states (b), a result of the boundary states taking measure from the itinerant states.  The boundary states are maximally localized at the bold cross.  The average transmission probability (c) ranges from $0$ for decoupled barriers (or edges) to $1$ for nonexistent barriers.}
	\label{f.5}
\end{figure}
Any quantum interference at the barrier must be destructive, which follows from $\Delta\rho_E^{}(0)=-(\rho_E^{}/4)(2-z_+^2-z_-^2)\le0$.  Because the undulations are generally not smooth at the barrier, the derivative $\Delta\rho_E^{\prime}(0^+)=(q\rho_E^{}/4)(z_+^3+z_-^3-z_+-z_-)$, shown in Fig.\,\ref{f.5}(a), is not necessarily positive.  In fact, as $\Delta\rho_E^{\prime}(0^+)$ depends only on odd powers of $z_\pm$, it is odd under the electron-hole operator $\mathcal{C}$: $\eta\mapsto-\eta$, and thus the signs of the derivative for electrons and holes are opposite.  Therefore, the undulations must exhibit a double minimum in half the parameter space [{\it cf.} the itinerant LDOS contribution in Fig.\,\ref{f.4}(b)].  Furthermore, these barriers could appear as two closely situated barriers.  It is an open question if this is related to the double-barrier features observed in graphene plasmon experiments.\cite{Fei13}

The undulations near the barrier could be masked by boundary states.  In this barrier model, there are $0$, $1$, or $2$ boundary states present.  Figure~\ref{f.5} shows that the number of boundary states correlates with $\Delta\rho_E^{\prime}(0^+)$ so that barriers with undulations exhibiting double-minima, also have boundary states.  This is a result of measure being transferred from the itinerant states to the boundary states, which can be shown in Fig.\,\ref{f.4}(b), where the sum of the LDOS contributions from itinerant and boundary states hovers around the graphene density of states.

The LDOS in Eq.\,(\ref{e.10}) and its undulations in Eq.\,(\ref{e.13}) and boundary contributions in Eq.\,(\ref{e.15}) can be fitted to experiment to extract the barrier parameters $\lambda$ and $\varepsilon$.  These parameters could then be inserted into Eq.\,(\ref{e.12}) to obtain the scattering properties, including the average transmission probability\cite{Gunl14}
\begin{equation}
	\bar{T} = \left\{
	\begin{array}{cc}
		\frac{\big.2\lambda^2}{\big.1-\varepsilon^2+\lambda^2}, & \quad|\eta\varepsilon\pm\lambda|<1,\\
		\frac{\big.\lambda}{\big.\eta\varepsilon+\lambda}, & \quad|\eta\varepsilon-\lambda|<1,~|\eta\varepsilon+\lambda|>1,\\
		-\frac{\big.\lambda}{\big.\eta\varepsilon-\lambda}, & \quad|\eta\varepsilon-\lambda|>1,~|\eta\varepsilon+\lambda|<1,\\
		\frac{\big.2\lambda^2}{\big(\varepsilon^2-\lambda^2\big)\big(\varepsilon^2-\lambda^2-1\big)}, & \quad|\eta\varepsilon\pm\lambda|>1,\\
	\end{array}
	\right.
	\label{e.17}
\end{equation}
plotted in Fig.\,\ref{f.5}(c).

For highly reflective barriers with weak coupling and weak potential, we could also use approximate expressions.  From the second order Born approximation, we obtain
\begin{equation}
	\Delta\rho_E^{}(x)\approx-\frac{\rho_E^{}}{2q|x|}\left[J_1+2\eta\varepsilon J_2-\big(\varepsilon^2+\lambda^2\big)\big(J_1-3J_3)\right].
	\label{e.18}
\end{equation}
The corresponding average transmission is $\bar T=2\lambda^2$.

\begin{figure}
    \includegraphics{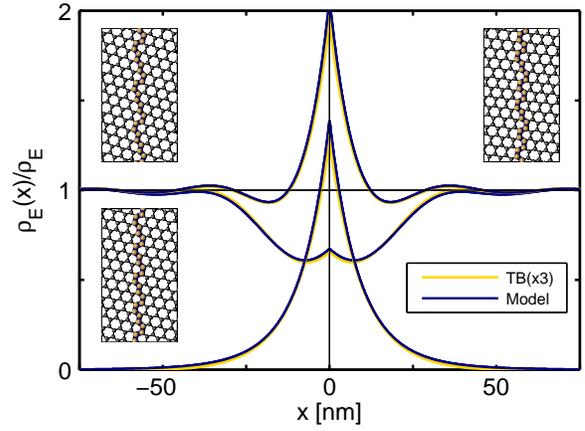}
    \caption{LDOS comparison between the graphene barrier model and the nearest-neighbor tight-binding (TB) model at the carrier energy $E=50$\,meV.  There are three sets of TB curves for the three barrier structures shown as insets, each with a different barrier orientation with respect to the graphene lattice.  The highlighted barrier sites have a potential $-1.2$\,eV.  The barrier model is computed for $(\lambda,\varepsilon)=(-1,-0.5)$.}
    \label{f.6}
\end{figure}
To test our analytical approach above, we have also calculated the LDOS numerically using an exact formalism within the tight-binding approximation.\cite{Lee81_1}  Figure~\ref{f.6} shows the results for three barriers with a constant potential along a chain of sites.  Although the graphene lattice orientations for the three barriers are quite different, the LDOS contributions are consistent and in excellent agreement with the barrier model.

In summary, we have derived the local density of states around a generic transport barrier described by an effective coupling parameter and an effective potential parameter.  We showed that the mesoscale undulations are related to the average transmission probability through the barrier, therefore making it possible to probe the transport properties through the barrier without the need for lateral transport measurements.  Rather, scanning probe techniques could be used to estimate the transmission probability.\cite{Clar13}  This could aid the search for suitable transport barriers in graphene, and down the road lead to new switchable graphene nano\-electronics.


\begin{acknowledgments}
The authors acknowledge support from the U.S. Office of Naval Research, directly and through the U.S. Naval Research Laboratory.
\end{acknowledgments}


%


\end{document}